\documentclass[journal]{IEEEtran}
\usepackage{amsfonts}
\usepackage{amsmath}
\usepackage{graphicx}
\usepackage{color}
\usepackage{multicol}
\usepackage{amssymb}

\begin{document}

\title{Outage probability of selective decode and forward relaying with secrecy constraints}

\IEEEoverridecommandlockouts
\author{Xiaojun Sun, Chunming Zhao,\\ 
\IEEEauthorblockA{National Mobile Communications Research Laboratory
(NCRL), \\Southeast University, Nanjing 210096, CHINA
\\Email: \{sunxiaojun, cmzhao2\}@seu.edu.cn}
}
 \maketitle

\begin{abstract}
We study the outage probability of opportunistic relay selection in
decode-and-forward relaying with secrecy constraints. We derive the
closed-form expression for the outage probability. Based on the
analytical result, the asymptotic performance is then investigated.
The accuracy of our performance analysis is verified by the
simulation results.
\end{abstract}


\section{Introduction}
Due to the broadcast nature of the transmission medium, wireless
communications cause serious security issues in practice.
Information-theoretic security has received much attention recently
[1]-[9]. The information-theoretic security was pioneered by Wyner
[1]. Later, the work in [2][3] extended Wyner's work to broadcast
channels and Gaussian channels, respectively.

Recently, the information-theoretic secure communications has been
generalized to wireless quasi-static fading channel [4]. The secure
multiple antennas system was also studied in [5]. However, multiple
antennas may not be available due to cost and size limitations.
Under this scenario, cooperative network is an efficient approach to
overcome this limitation. [6]-[9] discussed various relay or
cooperative strategies to increase security against eavesdroppers,
such as noise-forward [6], beam-forming [7], relay-jamming selection
[8]. The prior work in [9] considered relay selection for secure DF
cooperative communications. However, only the limiting value of the
outage probability is known [9], which is meaningful only at high
signal-to-noise ratios (SNRs). To the best of our knowledge, the
exact expression for the outage probability is still unknown in
selective decode-and-forward (DF) cooperation with secrecy
constraints.

In this letter, we study the outage probability of selective DF
cooperative secure communications over Rayleigh fading channels. As
the main contribution, we derive the analytical expressions for the
outage probability. Moreover, based on the analytical results, we
also investigate the asymptotic performance in the high SNR regime.
Simulation results verify the accuracy of our performance analysis.
\section{System model}
The half-duplex DF relay wireless system in Fig. 1 consists of one
source (S), $N$ trusted relays (R), one destination (D) and one
eavesdropper (E). Each node is equipped with single antenna.

\begin{figure}
\centering
\includegraphics[scale=0.5]{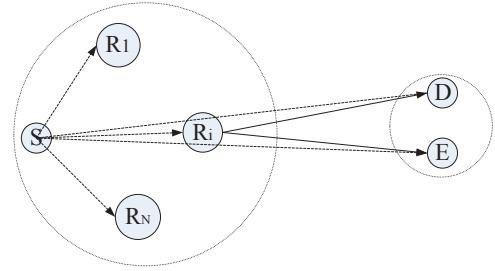}
\vspace{-0.3cm} \caption{Half-duplex DF relay wiretap channel model.
In the fist phase, S broadcasts the signal to all nodes (dotted
lines). In the second phase, the best relay node reforwards the
signal to D and E (solid lines).}
\end{figure}

The communication occurs in two hops. During the first hop, S
broadcasts the information to all nodes. For simplicity but without
loss of generality, we focus on the high SNR region where all the
relay nodes successfully decode the source transmission [7]-[9].
During the second hop, relay selection based on instantaneous
secrecy rate is performed. Let $\gamma _{sd} $, $\gamma _{se}$,
$\gamma _{rd}^n $ and $\gamma _{re}^n $ denote the instantaneous SNR
of the link $S \rightarrow D$, $S \rightarrow E$, $R_i \rightarrow
D$ and $R_i \rightarrow E$, respectively. All channels are subject
to Rayleigh fading. Thus, the PDFs of the SNRs, ${f\left( {\gamma
_{sd} } \right)}$, ${f\left( {\gamma _{se} } \right)}$, ${f\left(
{\gamma _{rd}^n } \right)}$ and ${f\left( {\gamma _{re}^n }
\right)}$, are exponentially distributed with parameter $\lambda
_{sd} $, $\lambda _{se} $, $\lambda _m $ and $\lambda _e$,
respectively.

\newcounter{mytempeqncnt}
\begin{figure*}[!t]
\normalsize \setcounter{mytempeqncnt}{\value{equation}}
\setcounter{equation}{7}
\begin{equation}
\begin{gathered}
  P_{out} \left( R \right) = 1 - \frac{{\lambda _{sd} }}
{{e^R \lambda _{se}  + \lambda _{sd} }}\exp \left( { - \frac{{e^R  -
1}} {{\lambda _{sd} }}} \right) + \left( {\frac{{e^R \lambda _e }}
{{e^R \lambda _e  + \lambda _m }}} \right)^N \frac{{\left( {\frac{1}
{{\lambda _{sd} }} + \frac{N} {{e^R \lambda _e }}} \right)^{ - 1} }}
{{e^R \lambda _{se}  + \lambda _{sd} }}\exp \left( { - \left( {e^R
- 1} \right)\left( {\frac{1} {{\lambda _{sd} }} + \frac{N} {{e^R
\lambda _e }} - \frac{N}
{{\lambda _m }}} \right)} \right) \hfill \\
  \;\;\;\;\;\;\;\;\;\;\;\;\; + \sum\limits_{n = 1}^N {C_n^N \left( {\frac{{ - \lambda _m }}
{{e^R \lambda _e  + \lambda _m }}} \right)^n } \exp \left( { -
\frac{{n\left( {e^R  - 1} \right)}} {{\lambda _m }}} \right)\frac{1}
{{e^R \lambda _{se}  + \lambda _{sd} }}\left[ {\frac{{e^R \lambda
_{se} \lambda _m }}
{{ne^R \lambda _{se}  + \lambda _m }} + f} \right], \hfill \\
\end{gathered}
\end{equation}
\setcounter{equation}{\value{mytempeqncnt}} \hrulefill \vspace*{4pt}
\end{figure*}

In the second hops, only the relay node with the largest
instantaneous secrecy rate is selected to forward the message to D.
Thus, D and E can combine the two received signals using maximal
ratio combining, respectively. Following the definition in [4], we
write the instantaneous secrecy rate about $n$th-relay link as
\begin{equation}
\begin{gathered}
  R_s^n= \max \left[ {\ln \left( {1 + \gamma _{m,n} } \right) - \ln \left( {1 + \gamma _{e,n} } \right),0} \right] \hfill \\
   = \max \left[ {\ln \left( {Z_n } \right),0} \right], \hfill \\
\end{gathered}
\end{equation}
where $\gamma _{m,n}={\gamma _{rd}^n  + \gamma _{sd} }$ denotes the
instantaneous SNR of the main channel, $\gamma _{e,n }={\gamma
_{re}^n  + \gamma _{se} }$ denotes the instantaneous SNR of the
eavesdropping channel, and $Z_n={1 + \gamma _{m,n} }/{1 + \gamma
_{e,n} }$. The output of the relay selection can be expressed as
\begin{equation}
Z_{\max }  = \max \left\{ {Z_1 , \cdots ,Z_N } \right\}
\end{equation}
with cumulative density function (CDF) as
\begin{equation}
F_{\max} \left( {z } \right) = \prod\limits_{n = 1}^N {F_n \left( {z
} \right)},
\end{equation}
where ${F_n \left( {z } \right)}$ is the CDF  of $Z_n$. After relay
selection, the instantaneous secrecy rate can be given by
\begin{equation}
\begin{gathered}
  R_s  = \max \left[ {\max \left( {\ln \left( {Z_n } \right)} \right),0} \right] = \max \left[ {\ln \left( {\max \left( {Z_n } \right)} \right),0} \right] \hfill \\
  \;\;\;\; = \max \left[ {\ln \left( {Z_{\max } } \right),0} \right]. \hfill \\
\end{gathered}
\end{equation}

This work characterizes the relay selection with secrecy constraints
in terms of outage probability as follows.

\section{Outage probability of relay selection with secrecy constraints}
Outage probability is an important performance measure, widely used
to characterize wireless communications. It is defined the
probability that the instantaneous secrecy capacity falls below a
target rate $R$ as
\begin{equation}
P_{out} \left( R \right) = P_r \left( {R_s \leqslant R} \right) =
F_{max} \left( {e^R } \right).
\end{equation}
\subsection{Relay selection without direct links}
In this subsection, we follow the system model in [9], where S has
no direct links with D and E.

In this case, the instantaneous SNR of the main channel and the
eavesdropping channel is $\gamma _{m,n}={\gamma _{rd}^n }$ and
$\gamma _{e,n }={\gamma _{re}^n }$, respectively. Therefore, the CDF
of $Z_n$, $F_n \left( z \right)$, can be expressed as
\begin{equation}
\begin{gathered}
  F_n \left( z \right) = \int_0^\infty  {f\left( {\gamma _{re}^n } \right)d\gamma _{re}^n \int_0^{z\gamma _{re}^n  + z - 1} {f\left( {\gamma _{rd}^n } \right)d\gamma _{rd}^n } }  \hfill \\
  \;\;\;\;\;\;\;\;\;\; = 1 - \exp \left( { - \frac{{z - 1}}
{{\lambda _m }}} \right)\frac{{\lambda _m }}
{{z\lambda _e  + \lambda _m }}. \hfill \\
\end{gathered}
\end{equation}
Substituting (6) in (3), $F_{\max} \left( {z } \right)$ can be
solved. Thus, in the case of independent identically distributed
(IID), using the binomial expansion, the outage probability for a
target rate $R$ is given by
\begin{equation}
P_{out} \left( R \right) = \sum\limits_{n = 0}^N {C_n^N \left(
{\frac{{ - \lambda _m }} {{e^R \lambda _e  + \lambda _m }}}
\right)^n \exp \left( { - \frac{{n\left( {e^R  - 1} \right)}}
{{\lambda _m }}} \right)},
\end{equation}
where $C_n^N  = N!/n!/\left( {N - n} \right)!$.
\subsection{Relay selection with direct links}
We extend the system model in [9] by considering the direct links
between S and D/E in this subsection.

In this scenario, the outage probability for a target rate $R$ (8)
is shown at the top of this page, where \setcounter{equation}{8}
\begin{equation}
f = \left\{ \begin{gathered}
  \frac{{\lambda _{sd} \lambda _m \left[ {\exp \left( {\left( {e^R  - 1} \right)\left( {\frac{n}
{{\lambda _m }} - \frac{1} {{\lambda _{sd} }}} \right)} \right) - 1}
\right]}}
{{n\lambda _{sd}  - \lambda _m }},\;\;n\lambda _{sd}  \ne \lambda _m  \hfill \\
  e^R  - 1,\;\;\;\;\;\;\;\;\;\;\;\;\;\;\;\;\;\;\;\;\;\;\;\;\;\;\;\;\;\;\;\;\;\;\;\;\;\;\;\;\;\;\;\;\;\;n\lambda _{sd}  = \lambda _m  \hfill \\
\end{gathered}  \right.
\end{equation}
\begin{IEEEproof}
The CDF of $Z_n$, ${F_n \left( {z } \right)}$, can be expressed as
\begin{equation}
F_n \left( z \right) = P_r \left( {\frac{{1 + \gamma _{rd}^n  +
\gamma _{sd} }} {{1 + \gamma _{re}^n  + \gamma _{se} }} < z} \right)
= P_r \left( {\gamma _{rd}^n < z\gamma _{re}^n  + u} \right),
\end{equation}
where $u = z\gamma _{se}  - \gamma _{sd} + z - 1$. The conditional
CDF $F_n \left( {z\left| u \right.} \right)$ is calculated as
follows.

In the case of $u \geqslant 0$, the conditional CDF $F_n \left(
{z\left| u \right.} \right)$ is given by
\begin{equation}
\begin{gathered}
  F_n \left( {z\left| u \right.} \right) = \int_0^\infty  {f\left( {\gamma _{re}^n } \right)d\gamma _{re}^n \int_0^{z\gamma _{re}^n  + u} {f\left( {\gamma _{rd}^n } \right)d\gamma _{rd}^n } }  \hfill \\
  \;\;\;\;\;\;\;\;\;\;\;\;\;\; = 1 - \exp \left( { - \frac{u}
{{\lambda _m }}} \right)\frac{{\lambda _m }}
{{z\lambda _e  + \lambda _m }} \hfill \\
\end{gathered}
\end{equation}
On the other hand, in the case of $u < 0$, the conditional CDF $F_n
\left( {z\left| u \right.} \right)$ can be expressed as
\begin{equation}
\begin{gathered}
  F_n \left( {z\left| u \right.} \right) = \int_{ - u/z}^\infty  {f\left( {\gamma _{re}^n } \right)d\gamma _{re}^n \int_0^{z\gamma _{re}^n  + u} {f\left( {\gamma _{rd}^n } \right)d\gamma _{rd}^n } }  \hfill \\
  \;\;\;\;\;\;\;\;\;\;\;\;\;\; = \exp \left( {\frac{u}
{{z\lambda _e }}} \right)\frac{{z\lambda _e }}
{{z\lambda _e  + \lambda _m }}. \hfill \\
\end{gathered}
\end{equation}
Therefore, using the binomial expansion, we can express the
conditional CDF $F_{\max} \left( {z\left| u \right.} \right)$
(maximum among $N$ IID random variable) as
\begin{equation}
F_{\max } \left( {z\left| u \right.} \right) = \left\{
\begin{gathered}
\sum\limits_{n = 0}^N {C_n^N \left( {\frac{{ - \lambda _m }}
{{z\lambda _e  + \lambda _m }}} \right)^n \exp \left( { -
\frac{{nu}}
{{\lambda _m }}} \right)} ,\;u \geqslant 0 \hfill \\
  \left( {\frac{{z\lambda _e }}
{{z\lambda _e  + \lambda _m }}} \right)^N \exp \left( {\frac{{Nu}}
{{z\lambda _e }}} \right),\;\;\;\;\;\;\;\;\;\;\;\;\;\;u < 0 \hfill \\
\end{gathered}  \right.
\end{equation}

Through introducing $v= z\gamma _{se} - \gamma _{sd}$, we can obtain
$F_{\max } \left( z \right)$ in the following. The PDF $f\left( v
\right)$ of $v$ is given by [10]
\begin{equation}
f\left( v \right) = \left\{ \begin{gathered}
  \frac{1}
{{z\lambda _{se}  + \lambda _{sd} }}\exp \left( { - \frac{v}
{{z\lambda _{se} }}} \right),\;\;\;v \geqslant 0 \hfill \\
  \frac{1}
{{z\lambda _{se}  + \lambda _{sd} }}\exp \left( {\frac{v}
{{\lambda _{sd} }}} \right),\;\;\;\;\;\;\;v < 0 \hfill \\
\end{gathered}  \right.
\end{equation}
Using (14), $u = v + z - 1$ and after simplifications, we can
express $P_{out} \left( R \right)$ as (8).
\end{IEEEproof}
\subsection{Asymptotic outage probability}
It is also important to examine the asymptotic behavior of the
outage probability at the high SNRs, where $\lambda _m  \to \infty$
and $\lambda _e  \to \infty $ with a constant $\kappa = \lambda _m
/\lambda _e $.

Without direct links, the asymptotic outage probability [9] is
expressed as
\begin{equation}
P_{out}^a \left( R \right) = \left( {\frac{{e^R }} {{e^R  + \kappa
}}} \right)^N.
\end{equation}

When S has direct links with D and E, we calculate the asymptotic
outage probability as follows. From (8), for fixed SNRs $\lambda
_{sd}$ and $\lambda _{se}$, we have
\begin{equation}
\begin{gathered}
  P_{out}^a \left( R \right) = 1 - \frac{{\lambda _{sd} }}
{{e^R \lambda _{se}  + \lambda _{sd} }}\exp \left( { - \frac{{e^R  -
1}}
{{\lambda _{sd} }}} \right) \hfill \\
  {\kern 1pt} {\kern 1pt} {\kern 1pt} {\kern 1pt} {\kern 1pt} {\kern 1pt} {\kern 1pt} {\kern 1pt} {\kern 1pt} {\kern 1pt} {\kern 1pt} {\kern 1pt} {\kern 1pt} {\kern 1pt} {\kern 1pt} {\kern 1pt}  + \left( {\frac{{e^R }}
{{e^R  + \kappa }}} \right)^N \frac{{e^R \lambda _{sd} }}
{{\left( {e^R \lambda _{se}  + \lambda _{sd} } \right)\left( {e^R  + \kappa _d } \right)}} \hfill \\
  {\kern 1pt} {\kern 1pt} {\kern 1pt} {\kern 1pt} {\kern 1pt} {\kern 1pt} {\kern 1pt} {\kern 1pt} {\kern 1pt} {\kern 1pt} {\kern 1pt} {\kern 1pt} {\kern 1pt} {\kern 1pt} {\kern 1pt} {\text{ + }}\sum\limits_{{\text{n = 1}}}^{\text{N}} {\frac{{{\text{C}}_{\text{n}}^{\text{N}} }}
{{e^R \lambda _{se}  + \lambda _{sd} }}} \left( {\frac{{ - \kappa }}
{{e^R  + \kappa }}} \right)^n \left[ {\frac{{e^R \lambda _{se} }}
{{ne^R \kappa _e  + 1}} + f_a }, \right] \hfill \\
\end{gathered}
\end{equation}
where ${\kappa _d =\lambda _{sd} /\lambda _e }$, ${\kappa _e=\lambda
_{se} /\lambda _m} $ and ${\kappa _m =\lambda _{sd} /\lambda _m }$.
If $n\kappa _m  - 1$, $f_a  = e^R  - 1$; else, $f_a  = \left( {\exp
\left( {\frac{{\left( {e^R  - 1} \right)\left( {n\kappa _m  - 1}
\right)}} {{\lambda _{sd} }}} \right) - 1} \right)\frac{{\lambda
_{sd} }} {{n\kappa _m  - 1}}$. In the case of  $\lambda _{sd}  \to
\infty$ and $\lambda _{se} \to \infty $ with a constant $\kappa_s =
\lambda _{sd} /\lambda _{se} $, (16) is re-expressed as
\begin{equation}
\begin{gathered}
  P_{out}^a \left( R \right) = \frac{{e^R }}
{{e^R  + \kappa _s }} + \left( {\frac{{e^R }} {{e^R  + \kappa }}}
\right)^N \frac{{e^R \kappa _s }}
{{\left( {e^R  + \kappa _s } \right)\left( {e^R  + N\kappa _d } \right)}} \hfill \\
  {\kern 1pt} {\kern 1pt} {\kern 1pt} {\kern 1pt} {\kern 1pt} {\kern 1pt} {\kern 1pt} {\kern 1pt} {\kern 1pt} {\kern 1pt} {\kern 1pt} {\kern 1pt} {\kern 1pt} {\kern 1pt} {\kern 1pt} {\text{ + }}\sum\limits_{{\text{n = 1}}}^{\text{N}} {\frac{{e^R {\text{C}}_{\text{n}}^{\text{N}} }}
{{\left( {e^R  + \kappa _s } \right)\left( {ne^R \kappa _e  + 1}
\right)}}\left( {\frac{{ - \kappa }}
{{e^R  + \kappa }}} \right)^n }.  \hfill \\
\end{gathered}
\end{equation}
If ${\kappa _d \to 0}$, ${\kappa _e \to 0} $ and ${\kappa _m \to
0}$, (16) and (17) can be re-expressed as (15). In this case, the
effect of direct links is negligible.
\begin{figure}
\centering
\includegraphics[scale=0.6]{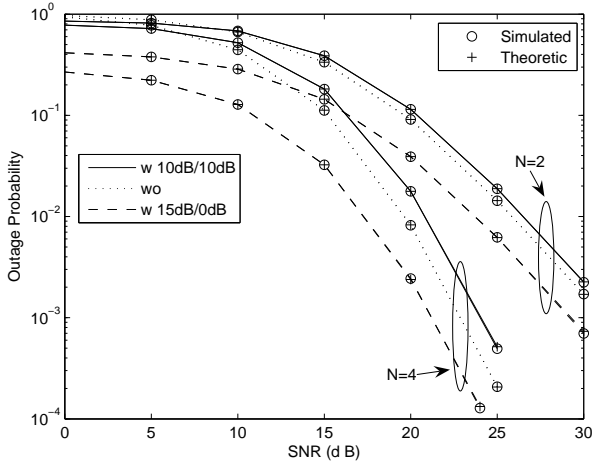}
\vspace{-0.3cm} \caption{Outage probability of $R=0.3$ V.S. $\lambda
_m$. 'wo' denotes that S has no direct links with D/E. 'w xdB/ydB'
denotes $\lambda _{sd}=xdB$ and $\lambda _{se}=ydB$.}
\end{figure}

\begin{figure}
\centering
\includegraphics[scale=0.6]{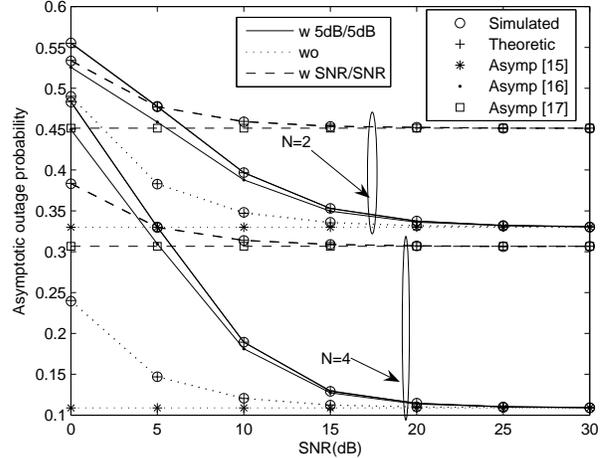}
\vspace{-0.3cm} \caption{Asymptotic outage probability of $R=0.3$
V.S. $\lambda _m$. 'wo' denotes that S has no direct links with D/E.
'w xdB/ydB' denotes $\lambda _{sd}=xdB$ and $\lambda _{se}=ydB$. 'w
SNR/SNR' denotes $\lambda _{sd}=\lambda _{se}=\lambda _m$.}
\end{figure}
\section{Simulation results}
Figure 2 shows the outage probability of $R=0.3$ for $\lambda
_e=15dB$ under different $\lambda _m$ and relay nodes $N$. It can be
observed that the experimental curves match exactly with the
theoretical results. These curves show that the outage probability
decreases with the increase of the number of relay nodes.

The asymptotic behavior of outage probability of $R=0.3$ as
functions of $\lambda _m$ under different relay nodes $N$ is
illustrated in Fig. 2. Here, SNR $\lambda _e$ is equal to SNR
$\lambda _m$. As can be seen from Fig. 2, these plotted curves
follow the above behavior. When ${\kappa _d, \kappa _e, \kappa _m
\to 0}$ with the increasing SNR $\lambda _m$, for example, SNRs
$\lambda _{sd}$ and $\lambda _{se}$ are fixed to be $5$dB, (16)
converges to (15) as shown in Fig. 2. Furthermore, the asymptotic
analysis efficiently converges to the true outage probability in the
high SNR regime. We also observe that the effect of direct links is
negligible in the high SNR regime when the ratios of SNRs, ${\kappa
_d}$, ${\kappa _e} $ and ${\kappa _m}$, are very small.

\section{Conclusion}
We have derived closed-form expression for the outage probability of
secure DF cooperative communications. It was shown that the relay
selection can reduce the outage probability. The experimental curves
are in excellent agreement with the theoretical results obtained in
this work. The future work will consider the relay-jamming
selection.

\end{document}